\documentclass[PRL, reprint,10pt,aps,amsmath,amssymb,longbibliography, superscriptaddress, nofootinbib]{revtex4-1}

\usepackage{graphicx}
\usepackage{dcolumn}
\usepackage{bm}
\usepackage{hyperref}
\usepackage{lipsum}
\usepackage{comment}
\usepackage{nameref}

\begin{document}

\title{Artificial intelligence moral agent as Adam Smith's impartial spectator}

\author{Nikodem Tomczak}
\email{tomczakn@imre.a-star.edu.sg}
\affiliation{Institute of Materials Research and Engineering, A*STAR (Agency for Science, Technology and Research), 2 Fusionopolis Way, Innovis, Singapore 138634}
\affiliation{Center on AI Technology for Humankind (AiTH), NUS Business School, National University of Singapore, 1 Business Link, Singapore 117592}

\begin{abstract}
Adam Smith developed a version of moral philosophy where better decisions are made by interrogating an impartial spectator within us. We discuss the possibility of using an external non-human-based substitute tool that would augment our internal mental processes and play the role of the impartial spectator. Such tool would have more knowledge about the world, be more impartial, and would provide a more encompassing 
perspective on moral assessment.
\end{abstract}

\maketitle

\section{Introduction}
Adam Smith developed a version of moral sentimentalism based on a strong commitment to the soundness of the ordinary human being's judgments based on common life experiences \cite{fleischacker2017}. He thought that individuals are much better off making their own moral decisions than any fixed system imposed either by intellectuals and governments or by natural or divine laws \cite{smithTMS}. However, Smith argued, to properly make normative judgements that would drive our decisions and minimize the corruption of our moral faculties one needs to consider the details of how we make decisions, and the details of the actions we judge, from an impartial spectator's perspective \cite{raphael2007}.

To interrogate an impartial spectator within ourselves is to undergo a cognitive process that makes sure that our views of what others feel in their circumstances match the feelings others actually have, and to avoid misjudging the situation out of ignorance or self-interest. Important in this context is Adam Smith's notion of sympathy related to the approbation of others of our actions - we should act to obtain sympathy \cite{rathbone2018}. In the context of business ethics, sympathy is crucial to avoid excessive self-interest and maintain societal harmony \cite{gonin2015adam}. Moral norms in this case express the feelings of an impartial spectator in relation to the decisions we make and how they affect others.

Some of the criticisms of Adam Smith's moral philosophy, in essence, are: (1) the impartial spectator will have biases representing interests that are grounded in the society it inhabits, (2) there is no clear procedure for how to interrogate the impartial spectator and decide which actions we should take, and (3) we can't tell upfront whether what the impartial spectator suggests is justified \cite{fleischacker2017}.

As our actions are based on the perceptions of others and adjusted to obtain sympathy, can we achieve the approbation of others by using an external non-human-based substitute tool that would augment our internal mental processes? As the impartial spectator is not supposed to be external and as it should also use sentiment rather than reason only as the basis of its judgments \cite{firth1952ethical} we may be inclined to say 'no'. External non-human agents, machines, from their virtue of being deterministic and designed and coded by humans, would not be able to do that completely on their own as genuine moral autonomy should, according to Kant, include a free will. Adam Smith acknowledges however that the impartial spectator we interrogate is not perfect but just a substitute representation of an ideal\footnote{Smith in \cite{smithTMS} refers also to a literal spectator who is "cool and impartial". It's the sum of the feelings of all such spectators that we should take into account. We take them into account by cultivating a perfect impartial spectator within us to watch over our conduct. We see ourselves through the eyes of other people.}. A substitution indeed, as a human being may strive to obtain but never really have perfect knowledge that would allow one to have perfect moral sentiments. This provides us with the possibility to introduce an artificial agent that would help to play the role of a spectator aspiring to impartiality.

Problems (1) and (2) are problems which artificial intelligence (AI) research is actively pursuing \cite{russell2016}. We may obtain broad AI models based on massive amounts of data, that are able to remove biases from ourselves and be impartial to the problems we try to solve \cite{simon1996the, nilsson2010the, goodfellow2016deep, parkes2015, hrnjic2019machine}. Current research on the ethics of AI-based decision-making and its influence on individuals and societies is trying to solve the problem (2) \cite{Dressel2018}. Problem (3) is much more complicated. Even if we understand the AI biases, and even if we know how conventional AI arrived at a decision, we still won't be able to tell whether we should or shouldn't take its advice into account \cite{mitchell2018prediction}. We could, however, reduce the impact of problem 3 by employing an artificial intelligence moral agent. AI moral agents could provide the impartial spectator with a more encompassing perspective on moral assessment. That is, we will provide ourselves with one more layer, albeit a dispassionate one, of assessment of our moral judgements and resulting decisions that could, in principle, improve in time. Overall, it seems therefore worthwhile to try developing artificial moral agents rather than adopt the consequentialist (utilitarian) approach \cite{card2020} in determining what actions we should take.

\section{Particularism and generalization in machine learning}

What transpires from Smith's approach to moral philosophy is that no fixed set of rules on how to make moral decisions for specific actions is provided but rather a framework for developing a virtuous character. Smith constructs the impartial spectator within us out of attitudes in the society around us. We seek the judgment of an impartial spectator within, rather than partial spectators without. Although the impartial spectator that resides within us connects us to society, it cannot however reach beyond the limits of the society without actually checking what lies beyond it. An important aspect of the impartial spectator approach in the context of decision-making is that we can hardly have universal general moral rules of right and wrong because these rules are derived from particular experiences, attitudes and interests, which may be different for different societies and individuals. This is the particularist view on how we construct moral judgement. What would be considered virtuous in one set of circumstances may not be so in a different set of circumstances. We can't have an impartial spectator within us that knows more than we do. In addition, humans have limited capability to experience and remember a multitude of circumstances within their lifetime so our impartial spectator may become deficient when a new set of circumstances presents itself \cite{forman2000adam}. 
 
A question arises then whether, and to what extent, we can outsource a fine-grained phenomenology of how we carry out various kinds of moral judgment in different circumstances to be free of society's biases \cite{zemel2013, Kleinberg2018b}. Can we build an external impartial spectator with which we could have an open dialogue equivalent to the one that is provided to us in our current circumstance to help us to care impartially for all human beings? 

The problem of heavy bias in our decisions, because of the anchoring of our views in our particular circumstance, could in principle be overcome by employing machine learning models that have been trained over much more heterogeneous population of social and ethical circumstances than those which are typically accessible to individuals, addressing to some extent problem (1).

However, if seeking the feelings of an impartial spectator is a source of moral norms, how the impartial spectator would know from which set of circumstances to choose from? An AI without knowledge outside of our society cannot be a meaningful impartial spectator. It may create a self-reinforcing loop where the deficiencies of our moral judgements will be amplified by AI suggestions. 

Ideally, such an impartial spectator would enable us to project ourselves into the lives of victims of injustice and thereby sympathize with them while correcting for systemic bias. In reality, we generally will build an impartial spectator within ourselves that will share the systemic biases found in the sentiments of our society. We know that bias in AI is a big largely unresolved that is often hard to correct for \cite{dwork2018, oneil2016}. Adopting an unsupervised particularist approach may just result in adopting particular biases related to particular circumstances rather than overcoming them. 

The advantage of having a perfect moral artificial agent would be to avoid failing in our ability to query the impartial spectator or it becoming a tool of our self-interest only, that is, that it doesn't become a simple mirror of our own initial assumptions but rather be free from partial feelings. This is particularly important when political and religious fanaticism perverts our moral feelings. Smith could not, however, as far as I know, find a clear solution for such a case.

We do not need to completely abandon or substitute our own impartial spectator for AI, but use it to complement our judgements in the case they present some deficiencies of which we are aware. If we recognize our own biases, we would at least be inclined to seek secondary advice. Instead of interrogating each member of a heterogeneous population, we may prefer to have an AI model that would generalize over all possible outcomes. In addition, AI is not supposed to provide normative judgements, the responsibility still stays with us; "every man is, no doubt, by nature, first and principally recommended to his care”, Smith says \citep{smithTMS}.

AI may provide a solution to the problem (2) because we know how interrogating an AI works. But knowing the workings is not enough - we often do not know how to interpret the results. Explainability of AI \cite{ribeiro2016, molnar2019} would be key to accepting it as any form of an impartial spectator because we need to be able to challenge it. If we can challenge and interrogate the outsourced impartial spectator in the same way as we would interrogate our own, and achieve a coherent outsourced expression of how an impartial spectator would feel about our conduct, there may still be no way for us to tell which source, the outside or the inside provides a more justified moral judgement. This would be still true despite the implemented particularism and our ability to explain the AI outcomes. 

This issue cannot, therefore, be solved with a simple AI agent alone. I see however two possible solutions to this problem. The first is to remove the absolute need for meeting the demands of an impartial spectator and fall back on consequentialism, departing therefore substantially from the foundations of Adam Smith's moral philosophy based on the propriety of our actions and not their consequences or utility. The second solution is to design an impartial spectator that is more than just an automated code for decision-making but truly provides solutions to human ethical and moral concerns.

Regarding the second solution, however, we didn't find a way, despite multiple attempts, to mechanize human moral values and moral reasoning, and associated emotions \cite{awad2018}. Moral reasoning may imply following a set of formal rules in a process that very much could be appeared as principled. Following moral rules is not however enough for an impartial spectator; it is exactly the trap of a particular set of rules that it tries to avoid - it needs also to understand emotions, without which there cannot be true sympathy. In human-machine interactions studies the problem of detecting emotions in humans and expressing human-readable emotions by machines is an old and very much contested research topic. We can't even agree on the number of non-verbally expressible emotions that humans have \cite{bartneck2020}. For the machines to understand justice and compassion and be programmed with ethical values, to be artificial moral agents, when making decisions they need to be much more than simple machine learning models evaluating possible outcomes of complex cases that are available today. 

\section{Quis custodiet ipsos custodes? Oversight and consequentialism}

Adam Smith's approach to how we decide which actions to take is often compared to that of consequentialism related to the writings of utilitarians \cite{lazari2017}, such as Jeremy Bentham \cite{bentham1970} or John Stuart Mill \cite{mill2006}. Where Smith paid more attention to the particularist conception of moral judgements and the propriety of actions and to motives that inspired them, the utilitarians paid more attention to the actions' outcomes and their consequences. This can be extended to the concept of fairness. While consequentialism will typically look at how our actions affect groups of people to maximize an overall measure of happiness, Smith in his moral evaluations pays more attention to the well-being of individuals, including those outside of our society. One important distinction between the two approaches is that for Adam Smith it implies we should seek to be informed by the views of people far outside our cultural communities, including those of disinterested people \citep{sen2009}. A structure of morality that reaches out across national and cultural borders is arguably extremely hard to come by for individuals without some external help. Consequentialism on the other hand does require to precisely define fairness, which is typically confined to much more narrowly defined problems and has a separate set of issues, as recently described by Card and Smith \cite{card2020}. In the context of choice, applying the impartial spectator approach avoids the problem of an objective way of characterizing the value of the outcome as such deliberations happen before the actual action - action is taken for the best outcome by definition. 

Human control of man-made machines is typically discussed in the context of the calamities they cause if allowed to function unattended. When machines fail, and they often do, they fail spectacularly in ways that were often hard to predict \cite{tenner1997}. So far AI cannot properly function without human oversight as it fails spectacularly too if anything out of the ordinary happens. We don't have yet self-driving cars without the need for human assistance, for good reasons, and probably won't have them for some time. One reason is that we can't absolve humans of ethical responsibility for the consequences of actions of systems that they have programmed to be autonomous. While humans have ethical commitments to others, machines don't. Accountability of individuals for their actions is part of the bargain that they have with their freedom. Any decision or action has a corresponding liability. 

In addition, when external shocks happen, such as the recent COVID-19 pandemic, the models trained on 'normal' data can't accommodate the 'new normal' data simply because they have never seen it but also because the predictions' time horizon was far more extended into the future than the rate at which the new circumstances were changing (this is an important point for decision making during the onset of and recovery from a crisis). For example, consumer behaviour changed in a way that could have not been anticipated by the AI-run recommendation engines. Similarly, sales, budget, and critical equipment forecasting broke down as customers shifted abruptly their priority to stocking essential products and disinfectants, while hospitals started to deplete extremely fast their standard PPE stock. From complex ethical decision-making to demand forecasting human intervention was unequivocally needed to prevent catastrophic failures \footnote{Whether a consequentialist or impartial spectator approach, or perhaps something else, would allow to minimize risk in a time of crisis is beyond the scope of this work.}. 

But there may be a partial solution to this. Instead of engineering an AI that provides moral guidance based on all possible circumstances, we should instead consider that the decision-making AI (perhaps even imbued with some type of consequentialism doctrine) could interrogate moral artificial intelligence agents within itself. These agents would play the role of impartial spectators forcing the main AI to act by moral demands. After Powers \cite{powers2009}, a system of moral agents solving one dedicated ethical problem out of the many that need to be considered for the final decision-making is reminiscent of a list of Kantian maxims of moral actions derived from Kant's basic moral law, the Categorical Imperative. It would be the AI's role then to recognize and understand how particular circumstances would map to a set of moral agents.

\section{Conclusions}
Adam Smith thought that moral philosophy can help us to make better decisions but can't and should not replace the common-life processes by which we make those decisions because moral decision-making is driven by emotion as much as by the intellect and is also shaped by our interactions with the people that may be affected by our actions. We need therefore to use deliberate thought grounded in ethics that corresponds to a particular situation. We do that by the process of interrogating an impartial spectator within us. The assessment of a particular situation however is typically a complex process for individuals with limited knowledge about the world. We could perhaps, at least partially, substitute or augment the impartial spectator with external artificial moral agents that would provide us with hints on how our actions are viewed by others. 

\section{Authors' Contributions}
NT conceived and planned the research, and wrote the manuscript.
\bibliography{bibliography.bib}

\begin{thebibliography}{30}%
\makeatletter
\providecommand \@ifxundefined [1]{%
 \@ifx{#1\undefined}
}%
\providecommand \@ifnum [1]{%
 \ifnum #1\expandafter \@firstoftwo
 \else \expandafter \@secondoftwo
 \fi
}%
\providecommand \@ifx [1]{%
 \ifx #1\expandafter \@firstoftwo
 \else \expandafter \@secondoftwo
 \fi
}%
\providecommand \natexlab [1]{#1}%
\providecommand \enquote  [1]{``#1''}%
\providecommand \bibnamefont  [1]{#1}%
\providecommand \bibfnamefont [1]{#1}%
\providecommand \citenamefont [1]{#1}%
\providecommand \href@noop [0]{\@secondoftwo}%
\providecommand \href [0]{\begingroup \@sanitize@url \@href}%
\providecommand \@href[1]{\@@startlink{#1}\@@href}%
\providecommand \@@href[1]{\endgroup#1\@@endlink}%
\providecommand \@sanitize@url [0]{\catcode `\\12\catcode `\$12\catcode
  `\&12\catcode `\#12\catcode `\^12\catcode `\_12\catcode `\%12\relax}%
\providecommand \@@startlink[1]{}%
\providecommand \@@endlink[0]{}%
\providecommand \url  [0]{\begingroup\@sanitize@url \@url }%
\providecommand \@url [1]{\endgroup\@href {#1}{\urlprefix }}%
\providecommand \urlprefix  [0]{URL }%
\providecommand \Eprint [0]{\href }%
\providecommand \doibase [0]{http://dx.doi.org/}%
\providecommand \selectlanguage [0]{\@gobble}%
\providecommand \bibinfo  [0]{\@secondoftwo}%
\providecommand \bibfield  [0]{\@secondoftwo}%
\providecommand \translation [1]{[#1]}%
\providecommand \BibitemOpen [0]{}%
\providecommand \bibitemStop [0]{}%
\providecommand \bibitemNoStop [0]{.\EOS\space}%
\providecommand \EOS [0]{\spacefactor3000\relax}%
\providecommand \BibitemShut  [1]{\csname bibitem#1\endcsname}%
\let\auto@bib@innerbib\@empty
\bibitem [{\citenamefont {Fleischacker}(2017)}]{fleischacker2017}%
  \BibitemOpen
  \bibfield  {author} {\bibinfo {author} {\bibfnamefont {Samuel}\ \bibnamefont
  {Fleischacker}},\ }\bibfield  {title} {\enquote {\bibinfo {title} {Adam
  smith's moral and political philosophy},}\ }in\ \href@noop {} {\emph
  {\bibinfo {booktitle} {The Stanford Encyclopedia of Philosophy}}},\ \bibinfo
  {editor} {edited by\ \bibinfo {editor} {\bibfnamefont {Edward~N.}\
  \bibnamefont {Zalta}}}\ (\bibinfo  {publisher} {Metaphysics Research Lab,
  Stanford University},\ \bibinfo {year} {2017})\ \bibinfo {edition} {spring
  2017}\ ed.\BibitemShut {Stop}%
\bibitem [{\citenamefont {Smith}\ \emph {et~al.}(2009)\citenamefont {Smith},
  \citenamefont {Hanley},\ and\ \citenamefont {Sen}}]{smithTMS}%
  \BibitemOpen
  \bibfield  {author} {\bibinfo {author} {\bibfnamefont {Adam}\ \bibnamefont
  {Smith}}, \bibinfo {author} {\bibfnamefont {Ryan~Patrick}\ \bibnamefont
  {Hanley}}, \ and\ \bibinfo {author} {\bibfnamefont {Amartya}\ \bibnamefont
  {Sen}},\ }\href@noop {} {\emph {\bibinfo {title} {The Theory of Moral
  Sentiments}}}\ (\bibinfo  {publisher} {Penguin Classics},\ \bibinfo {year}
  {2009})\BibitemShut {NoStop}%
\bibitem [{\citenamefont {Raphael}(2007)}]{raphael2007}%
  \BibitemOpen
  \bibfield  {author} {\bibinfo {author} {\bibfnamefont {D~D}\ \bibnamefont
  {Raphael}},\ }\href@noop {} {\emph {\bibinfo {title} {The Impartial
  Spectator: Adam Smith's Moral Philosophy}}}\ (\bibinfo  {publisher} {Oxford
  University Press, USA},\ \bibinfo {year} {2007})\BibitemShut {NoStop}%
\bibitem [{\citenamefont {Rathbone}(2018)}]{rathbone2018}%
  \BibitemOpen
  \bibfield  {author} {\bibinfo {author} {\bibfnamefont {Mark}\ \bibnamefont
  {Rathbone}},\ }\bibfield  {title} {\enquote {\bibinfo {title} {Adam smith,
  the impartial spectator and embodiment: Towards an economics of
  accountability and dialogue},}\ }\href {\doibase 10.3390/rel9040118}
  {\bibfield  {journal} {\bibinfo  {journal} {Religions}\ }\textbf {\bibinfo
  {volume} {9}} (\bibinfo {year} {2018}),\ 10.3390/rel9040118}\BibitemShut
  {NoStop}%
\bibitem [{\citenamefont {Gonin}(2015)}]{gonin2015adam}%
  \BibitemOpen
  \bibfield  {author} {\bibinfo {author} {\bibfnamefont {Michael}\ \bibnamefont
  {Gonin}},\ }\bibfield  {title} {\enquote {\bibinfo {title} {Adam smith’s
  contribution to business ethics, then and now},}\ }\href@noop {} {\bibfield
  {journal} {\bibinfo  {journal} {Journal of Business Ethics}\ }\textbf
  {\bibinfo {volume} {129}},\ \bibinfo {pages} {221--236} (\bibinfo {year}
  {2015})}\BibitemShut {NoStop}%
\bibitem [{\citenamefont {Firth}(1952)}]{firth1952ethical}%
  \BibitemOpen
  \bibfield  {author} {\bibinfo {author} {\bibfnamefont {Roderick}\
  \bibnamefont {Firth}},\ }\bibfield  {title} {\enquote {\bibinfo {title}
  {Ethical absolutism and the ideal observer},}\ }\href@noop {} {\bibfield
  {journal} {\bibinfo  {journal} {Philosophy and Phenomenological Research}\
  }\textbf {\bibinfo {volume} {12}},\ \bibinfo {pages} {317--345} (\bibinfo
  {year} {1952})}\BibitemShut {NoStop}%
\bibitem [{\citenamefont {Russell}\ and\ \citenamefont
  {Norvig}(2016)}]{russell2016}%
  \BibitemOpen
  \bibfield  {author} {\bibinfo {author} {\bibfnamefont {Stuart}\ \bibnamefont
  {Russell}}\ and\ \bibinfo {author} {\bibfnamefont {Peter}\ \bibnamefont
  {Norvig}},\ }\href@noop {} {\emph {\bibinfo {title} {Artificial intelligence:
  a modern approach}}}\ (\bibinfo  {publisher} {Pearson; 3 edition},\ \bibinfo
  {year} {2016})\BibitemShut {NoStop}%
\bibitem [{\citenamefont {Simon}(1996)}]{simon1996the}%
  \BibitemOpen
  \bibfield  {author} {\bibinfo {author} {\bibfnamefont {Herbert}\ \bibnamefont
  {Simon}},\ }\href@noop {} {\emph {\bibinfo {title} {The sciences of the
  artificial, 3rd ed.}}}\ (\bibinfo  {publisher} {MIT Press},\ \bibinfo
  {address} {Cambridge, Mass},\ \bibinfo {year} {1996})\BibitemShut {NoStop}%
\bibitem [{\citenamefont {Nilsson}(2010)}]{nilsson2010the}%
  \BibitemOpen
  \bibfield  {author} {\bibinfo {author} {\bibfnamefont {Nils}\ \bibnamefont
  {Nilsson}},\ }\href@noop {} {\emph {\bibinfo {title} {The quest for
  artificial intelligence : a history of ideas and achievements}}}\ (\bibinfo
  {publisher} {Cambridge University Press},\ \bibinfo {address} {Cambridge New
  York},\ \bibinfo {year} {2010})\BibitemShut {NoStop}%
\bibitem [{\citenamefont {Goodfellow}(2016)}]{goodfellow2016deep}%
  \BibitemOpen
  \bibfield  {author} {\bibinfo {author} {\bibfnamefont {Ian}\ \bibnamefont
  {Goodfellow}},\ }\href@noop {} {\emph {\bibinfo {title} {Deep learning}}}\
  (\bibinfo  {publisher} {The MIT Press},\ \bibinfo {address} {Cambridge,
  Massachusetts},\ \bibinfo {year} {2016})\BibitemShut {NoStop}%
\bibitem [{\citenamefont {Parkes}\ and\ \citenamefont
  {Wellman}(2015)}]{parkes2015}%
  \BibitemOpen
  \bibfield  {author} {\bibinfo {author} {\bibfnamefont {David~C}\ \bibnamefont
  {Parkes}}\ and\ \bibinfo {author} {\bibfnamefont {Michael~P}\ \bibnamefont
  {Wellman}},\ }\bibfield  {title} {\enquote {\bibinfo {title} {Economic
  reasoning and artificial intelligence},}\ }\href@noop {} {\bibfield
  {journal} {\bibinfo  {journal} {Science}\ }\textbf {\bibinfo {volume}
  {349}},\ \bibinfo {pages} {267--272} (\bibinfo {year} {2015})}\BibitemShut
  {NoStop}%
\bibitem [{\citenamefont {Hrnjic}\ and\ \citenamefont
  {Tomczak}(2019)}]{hrnjic2019machine}%
  \BibitemOpen
  \bibfield  {author} {\bibinfo {author} {\bibfnamefont {Emir}\ \bibnamefont
  {Hrnjic}}\ and\ \bibinfo {author} {\bibfnamefont {Nikodem}\ \bibnamefont
  {Tomczak}},\ }\bibfield  {title} {\enquote {\bibinfo {title} {Machine
  learning and behavioral economics for personalized choice architecture},}\
  }\href@noop {} {\bibfield  {journal} {\bibinfo  {journal} {arXiv preprint
  arXiv:1907.02100}\ } (\bibinfo {year} {2019})}\BibitemShut {NoStop}%
\bibitem [{\citenamefont {Dressel}\ and\ \citenamefont
  {Farid}(2018)}]{Dressel2018}%
  \BibitemOpen
  \bibfield  {author} {\bibinfo {author} {\bibfnamefont {Julia}\ \bibnamefont
  {Dressel}}\ and\ \bibinfo {author} {\bibfnamefont {Hany}\ \bibnamefont
  {Farid}},\ }\bibfield  {title} {\enquote {\bibinfo {title} {The accuracy,
  fairness, and limits of predicting recidivism},}\ }\href@noop {} {\bibfield
  {journal} {\bibinfo  {journal} {Science Advances}\ }\textbf {\bibinfo
  {volume} {4}} (\bibinfo {year} {2018})}\BibitemShut {NoStop}%
\bibitem [{\citenamefont {Mitchell}\ \emph {et~al.}(2018)\citenamefont
  {Mitchell}, \citenamefont {Potash},\ and\ \citenamefont
  {Barocas}}]{mitchell2018prediction}%
  \BibitemOpen
  \bibfield  {author} {\bibinfo {author} {\bibfnamefont {Shira}\ \bibnamefont
  {Mitchell}}, \bibinfo {author} {\bibfnamefont {Eric}\ \bibnamefont {Potash}},
  \ and\ \bibinfo {author} {\bibfnamefont {Solon}\ \bibnamefont {Barocas}},\
  }\bibfield  {title} {\enquote {\bibinfo {title} {Prediction-based decisions
  and fairness: A catalogue of choices, assumptions, and definitions},}\
  }\href@noop {} {\bibfield  {journal} {\bibinfo  {journal} {arXiv preprint}\
  ,\ \bibinfo {eid} {arXiv:1811.07867}} (\bibinfo {year} {2018})}\BibitemShut
  {NoStop}%
\bibitem [{\citenamefont {Card}\ and\ \citenamefont {A}(2020)}]{card2020}%
  \BibitemOpen
  \bibfield  {author} {\bibinfo {author} {\bibfnamefont {Dallas}\ \bibnamefont
  {Card}}\ and\ \bibinfo {author} {\bibfnamefont {Smith~Noah}\ \bibnamefont
  {A}},\ }\bibfield  {title} {\enquote {\bibinfo {title} {{On consequentialism
  and fairness}},}\ }\href@noop {} {\bibfield  {journal} {\bibinfo  {journal}
  {arXiv e-prints}\ ,\ \bibinfo {eid} {arXiv:2001.00329v2}} (\bibinfo {year}
  {2020})}\BibitemShut {NoStop}%
\bibitem [{\citenamefont {Forman-Barzilai}(2000)}]{forman2000adam}%
  \BibitemOpen
  \bibfield  {author} {\bibinfo {author} {\bibfnamefont {Fonna}\ \bibnamefont
  {Forman-Barzilai}},\ }\bibfield  {title} {\enquote {\bibinfo {title} {Adam
  smith as globalization theorist},}\ }\href@noop {} {\bibfield  {journal}
  {\bibinfo  {journal} {Critical Review}\ }\textbf {\bibinfo {volume} {14}},\
  \bibinfo {pages} {391--419} (\bibinfo {year} {2000})}\BibitemShut {NoStop}%
\bibitem [{\citenamefont {Zemel}\ \emph {et~al.}(2013)\citenamefont {Zemel},
  \citenamefont {Wu}, \citenamefont {Swersky}, \citenamefont {Pitassi},\ and\
  \citenamefont {Dwork}}]{zemel2013}%
  \BibitemOpen
  \bibfield  {author} {\bibinfo {author} {\bibfnamefont {Rich}\ \bibnamefont
  {Zemel}}, \bibinfo {author} {\bibfnamefont {Yu}~\bibnamefont {Wu}}, \bibinfo
  {author} {\bibfnamefont {Kevin}\ \bibnamefont {Swersky}}, \bibinfo {author}
  {\bibfnamefont {Toni}\ \bibnamefont {Pitassi}}, \ and\ \bibinfo {author}
  {\bibfnamefont {Cynthia}\ \bibnamefont {Dwork}},\ }\bibfield  {title}
  {\enquote {\bibinfo {title} {Learning fair representations},}\ }\href@noop {}
  {\bibfield  {journal} {\bibinfo  {journal} {Proceedings of International
  Conference on Machine Learning}\ ,\ \bibinfo {pages} {325--333}} (\bibinfo
  {year} {2013})}\BibitemShut {NoStop}%
\bibitem [{\citenamefont {Kleinberg}\ \emph {et~al.}(2018)\citenamefont
  {Kleinberg}, \citenamefont {Ludwig}, \citenamefont {Mullainathan},\ and\
  \citenamefont {Rambachan}}]{Kleinberg2018b}%
  \BibitemOpen
  \bibfield  {author} {\bibinfo {author} {\bibfnamefont {Jon}\ \bibnamefont
  {Kleinberg}}, \bibinfo {author} {\bibfnamefont {Jens}\ \bibnamefont
  {Ludwig}}, \bibinfo {author} {\bibfnamefont {Sendhil}\ \bibnamefont
  {Mullainathan}}, \ and\ \bibinfo {author} {\bibfnamefont {Ashesh}\
  \bibnamefont {Rambachan}},\ }\bibfield  {title} {\enquote {\bibinfo {title}
  {Algorithmic fairness},}\ }\href@noop {} {\bibfield  {journal} {\bibinfo
  {journal} {AEA Papers and Proceedings}\ }\textbf {\bibinfo {volume} {108}},\
  \bibinfo {pages} {22--27} (\bibinfo {year} {2018})}\BibitemShut {NoStop}%
\bibitem [{\citenamefont {Dwork}\ \emph {et~al.}(2018)\citenamefont {Dwork},
  \citenamefont {Immorlica}, \citenamefont {Tauman~Kalai},\ and\ \citenamefont
  {Leiserson}}]{dwork2018}%
  \BibitemOpen
  \bibfield  {author} {\bibinfo {author} {\bibfnamefont {Cynthia}\ \bibnamefont
  {Dwork}}, \bibinfo {author} {\bibfnamefont {Nicole}\ \bibnamefont
  {Immorlica}}, \bibinfo {author} {\bibfnamefont {Adam}\ \bibnamefont
  {Tauman~Kalai}}, \ and\ \bibinfo {author} {\bibfnamefont {Max}\ \bibnamefont
  {Leiserson}},\ }\bibfield  {title} {\enquote {\bibinfo {title} {Decoupled
  classifiers for group-fair and efficient machine learning},}\ }\href@noop {}
  {\bibfield  {journal} {\bibinfo  {journal} {Proceedings of the 1st Conference
  on Fairness, Accountability and Transparency}\ }\bibinfo {series}
  {Proceedings of Machine Learning Research},\ \textbf {\bibinfo {volume}
  {81}},\ \bibinfo {pages} {119--133} (\bibinfo {year} {2018})}\BibitemShut
  {NoStop}%
\bibitem [{\citenamefont {O'Neil}(2016)}]{oneil2016}%
  \BibitemOpen
  \bibfield  {author} {\bibinfo {author} {\bibfnamefont {Cathy}\ \bibnamefont
  {O'Neil}},\ }\href@noop {} {\emph {\bibinfo {title} {Weapons of Math
  Destruction: How Big Data Increases Inequality and Threatens Democracy}}}\
  (\bibinfo  {publisher} {Broadway Books},\ \bibinfo {year} {2016})\BibitemShut
  {NoStop}%
\bibitem [{\citenamefont {Ribeiro}\ \emph {et~al.}(2016)\citenamefont
  {Ribeiro}, \citenamefont {Singh},\ and\ \citenamefont
  {Guestrin}}]{ribeiro2016}%
  \BibitemOpen
  \bibfield  {author} {\bibinfo {author} {\bibfnamefont {Marco~Tulio}\
  \bibnamefont {Ribeiro}}, \bibinfo {author} {\bibfnamefont {Sameer}\
  \bibnamefont {Singh}}, \ and\ \bibinfo {author} {\bibfnamefont {Carlos}\
  \bibnamefont {Guestrin}},\ }\href@noop {} {\enquote {\bibinfo {title} {{''Why
  Should I Trust You?'': Explaining the Predictions of Any Classifier}},}\ }
  (\bibinfo {year} {2016})\BibitemShut {NoStop}%
\bibitem [{\citenamefont {Molnar}(2019)}]{molnar2019}%
  \BibitemOpen
  \bibfield  {author} {\bibinfo {author} {\bibfnamefont {Christoph}\
  \bibnamefont {Molnar}},\ }\href@noop {} {\emph {\bibinfo {title}
  {{Interpretable Machine Learning: A Guide for Making Black Box Models
  Explainable}}}}\ (\bibinfo  {publisher} {Christoph Molnar},\ \bibinfo {year}
  {2019})\ \bibinfo {note} {accessed: 2019-05-30}\BibitemShut {NoStop}%
\bibitem [{\citenamefont {Awad}\ \emph {et~al.}(2018)\citenamefont {Awad},
  \citenamefont {Dsouza}, \citenamefont {Kim}, \citenamefont {Schulz},
  \citenamefont {Henrich}, \citenamefont {Shariff}, \citenamefont {Bonnefon},\
  and\ \citenamefont {Rahwan}}]{awad2018}%
  \BibitemOpen
  \bibfield  {author} {\bibinfo {author} {\bibfnamefont {Edmond}\ \bibnamefont
  {Awad}}, \bibinfo {author} {\bibfnamefont {Sohan}\ \bibnamefont {Dsouza}},
  \bibinfo {author} {\bibfnamefont {Richard}\ \bibnamefont {Kim}}, \bibinfo
  {author} {\bibfnamefont {Jonathan}\ \bibnamefont {Schulz}}, \bibinfo {author}
  {\bibfnamefont {Joseph}\ \bibnamefont {Henrich}}, \bibinfo {author}
  {\bibfnamefont {Azim}\ \bibnamefont {Shariff}}, \bibinfo {author}
  {\bibfnamefont {Jean-Fran{\c{c}}ois}\ \bibnamefont {Bonnefon}}, \ and\
  \bibinfo {author} {\bibfnamefont {Iyad}\ \bibnamefont {Rahwan}},\ }\bibfield
  {title} {\enquote {\bibinfo {title} {The moral machine experiment},}\
  }\href@noop {} {\bibfield  {journal} {\bibinfo  {journal} {Nature}\ }\textbf
  {\bibinfo {volume} {563}},\ \bibinfo {pages} {59} (\bibinfo {year}
  {2018})}\BibitemShut {NoStop}%
\bibitem [{\citenamefont {Bartneck}\ \emph {et~al.}(2020)\citenamefont
  {Bartneck}, \citenamefont {Belpaeme}, \citenamefont {Eyssel}, \citenamefont
  {Kanda}, \citenamefont {Keijsers},\ and\ \citenamefont
  {Sabanovic}}]{bartneck2020}%
  \BibitemOpen
  \bibfield  {author} {\bibinfo {author} {\bibfnamefont {Christoph}\
  \bibnamefont {Bartneck}}, \bibinfo {author} {\bibfnamefont {Tony}\
  \bibnamefont {Belpaeme}}, \bibinfo {author} {\bibfnamefont {Friederike}\
  \bibnamefont {Eyssel}}, \bibinfo {author} {\bibfnamefont {Takayuki}\
  \bibnamefont {Kanda}}, \bibinfo {author} {\bibfnamefont {Merel}\ \bibnamefont
  {Keijsers}}, \ and\ \bibinfo {author} {\bibfnamefont {Selma}\ \bibnamefont
  {Sabanovic}},\ }\href@noop {} {\emph {\bibinfo {title} {Human-Robot
  Interaction: An Introduction}}}\ (\bibinfo  {publisher} {Cambridge University
  Press},\ \bibinfo {year} {2020})\BibitemShut {NoStop}%
\bibitem [{\citenamefont {de~Lazari-Radek}\ and\ \citenamefont
  {Singer}(2017)}]{lazari2017}%
  \BibitemOpen
  \bibfield  {author} {\bibinfo {author} {\bibfnamefont {Katarzyna}\
  \bibnamefont {de~Lazari-Radek}}\ and\ \bibinfo {author} {\bibfnamefont
  {Peter}\ \bibnamefont {Singer}},\ }\href@noop {} {\emph {\bibinfo {title}
  {Utilitarianism. A very short introduction}}}\ (\bibinfo  {publisher} {Oxford
  University Press},\ \bibinfo {year} {2017})\BibitemShut {NoStop}%
\bibitem [{\citenamefont {Bentham}(1970)}]{bentham1970}%
  \BibitemOpen
  \bibfield  {author} {\bibinfo {author} {\bibfnamefont {Jeremy}\ \bibnamefont
  {Bentham}},\ }\href@noop {} {\emph {\bibinfo {title} {An Introduction to the
  Principles of Morals and Legislation}}},\ [1781]\ (\bibinfo  {publisher}
  {Oxford University Press},\ \bibinfo {year} {1970})\BibitemShut {NoStop}%
\bibitem [{\citenamefont {Mill}(2006)}]{mill2006}%
  \BibitemOpen
  \bibfield  {author} {\bibinfo {author} {\bibfnamefont {John~Stuart}\
  \bibnamefont {Mill}},\ }\href@noop {} {\emph {\bibinfo {title} {Essays on
  Ethics, Religion and Society (Utilitarianism) [1833]}}},\ Collected Works of
  John Stuart Mill 10\ (\bibinfo  {publisher} {Liberty Fund},\ \bibinfo {year}
  {2006})\BibitemShut {NoStop}%
\bibitem [{\citenamefont {Sen}(2009)}]{sen2009}%
  \BibitemOpen
  \bibfield  {author} {\bibinfo {author} {\bibfnamefont {Amartya}\ \bibnamefont
  {Sen}},\ }\href@noop {} {\emph {\bibinfo {title} {The Idea of Justice}}}\
  (\bibinfo  {publisher} {Belknap Press},\ \bibinfo {year} {2009})\BibitemShut
  {NoStop}%
\bibitem [{\citenamefont {Tenner}(1997)}]{tenner1997}%
  \BibitemOpen
  \bibfield  {author} {\bibinfo {author} {\bibfnamefont {Edward}\ \bibnamefont
  {Tenner}},\ }\href@noop {} {\emph {\bibinfo {title} {Why Things Bite Back:
  Technology and the Revenge of Unintended Consequences}}}\ (\bibinfo
  {publisher} {Vintage},\ \bibinfo {year} {1997})\BibitemShut {NoStop}%
\bibitem [{\citenamefont {Powers}(20009)}]{powers2009}%
  \BibitemOpen
  \bibfield  {author} {\bibinfo {author} {\bibfnamefont {Thomas~M}\
  \bibnamefont {Powers}},\ }\bibfield  {title} {\enquote {\bibinfo {title}
  {Machines and moral reasoning},}\ }\href@noop {} {\bibfield  {journal}
  {\bibinfo  {journal} {Philosophy Now}\ }\textbf {\bibinfo {volume} {72}}
  (\bibinfo {year} {20009})}\BibitemShut {NoStop}%
\end{thebibliography}%
\end{document}